# A portable code for dissipative particle dynamics (DPD) simulations with additional specific interactions

Hideo Doi [a,#,$], Koji Okuwaki [a], Takamitsu Naito [a], Sona Saitou [a],

Yuji Mochizuki [a,b,*]

[a] *Department of Chemistry and Research Center for Smart Molecules, Faculty of Science, Rikkyo University, 3-34-1 Nishi-ikebukuro, Toshima-ku, Tokyo 171-8501, Japan*

[b] *Institute of Industrial Science, The University of Tokyo, 4-6-1 Komaba, Meguro-ku, Tokyo 153-8503, Japan*

\* 1st corresponding author: fullmoon@rikkyo.ac.jp (Yuji Mochizuki)

\# 2nd corresponding author: hideo-doi@rikkyo.ac.jp (Hideo Doi)

($ Present address: Research Center for Computational Design of Advanced Functional Materials, AIST Tsukuba Central 2, 1-1-1 Umezono, Tsukuba, Ibaraki 305-8568, Japan / Email: hideo-doi@aist.go.jp)

# Abstract

We developed a portable code for dissipative particle dynamics (DPD) simulations. This Fortran program named CAMUS has a couple of notable features. One is the omission of constructing the so-called neighboring particles list, providing a sizable speed-up per step and also a near linear scaling of costs with respect to the number of particles. The other is an easy inclusion of additional specific (such as 1-3 and 1-5 Morse bonding) interactions which are crucial in describing protein structures. The formations of $\alpha$-helix and $\beta$-sheet through DPD were then demonstrated. CAMUS is freely available at the GitHub site.

# Introduction

There have always been demands to predict and investigate molecular properties, hence various methods and related programs have been developed for conducting such a research area. Particularly, the method of dissipative particle dynamics (DPD) has attracted attentions in recent years. In DPD simulations, atoms or molecules are generally treated as coarse-grained particles, and the number of interactions to be computed in a given system is greatly reduced through the so-called soft potentials (as addressed later) [1, 2]. Thus, DPD simulations of large-scale systems are relatively feasible, e.g. even for membranes [3, 4]. DPD has a merit to take long effective time steps in comparison with the method of coarse-grained molecular dynamics (CG-MD) [5], and thus long time evolutions of molecules become tractable. The treatment of highly directional specific interactions such as hydrogen bonding has been difficult in conventional DPD simulations, however.

COGNAC [6] is a DPD code that has been developed as a module of the OCTA program suite for soft materials [7–9]. COGNAC provides general-purpose MD simulations covering from atomistic molecular models to coarse-grained models: various applications with COGNAC have been reported such as in Refs. [10–12]. In 2012, Vishnyakov et al.

reported that the protein structures of α-helix and β-hairpin are successfully modeled by DPD simulations [13]. Note that the local and site-directional hydrogen bonding plays crucial roles in forming such specific structures of proteins [14]. In Ref. [13], the 1-3 and 1-5 interaction potentials of the Morse type were employed to effectively model hydrogen bonds. Unfortunately, the use of such additional potentials is rather difficult with COGNAC. Other major DPD-usable codes, LAMMPS [15, 16], HOOMD-blue [17–19], GROMACS [20, 21], and DL_MESO [22] have been well-matured, and the complexity and largeness of these program systems would rather restrict the style to introduce new potentials or to modify related functionalities. Thus, we have decided to develop an original DPD code to which various modifications and extensions are easily applied.

In this paper, we report the development of a portable DPD code. This Fortran program (about 3000 lines) is named CAMUS (Code for dissipative particle dynAMics simUlationS). CAMUS has a flexibility to concisely introduce additional potentials describing local and directional interactions needed for such as hydrogen bonding [13]. The remaining parts of this paper are composed as follows. After summarizing the DPD simulation, the design concept of CAMUS is described. In the section of results and discussion, the tests of parallel performance are shown, and then the DPD simulations of protein structure are demonstrated.

## Summary of DPD simulation

DPD is based on the dynamics of soft particles interacting by conservative, dissipative, and random forces [23, 24]. The fundamental DPD scheme was extended to polymer system by Groot et al., by introducing a bead-spring type particle model [2, 25, 26].

Here, Groot's DPD model for polymers [2, 25, 26] is outlined as follows. The time evolution of the given system under forces $\mathbf{f}_i$ is simulated by solving the standard Newtonian equation of motion

$$\frac{d\mathbf{r}_i}{dt} = \mathbf{v}_i \tag{1}$$

and

$$m_i \frac{d\mathbf{v}_i}{dt} = \mathbf{f}_i, \tag{2}$$

where $\mathbf{r}_i$, $\mathbf{v}_i$, and $m_i$ are the position vector, velocity vector, and mass of the $i$th particle, respectively. The masses and diameters of particles are made dimensionless. There have been several schemes of actual time evolution (or integration) [21, 25]

As mentioned above, the force $\mathbf{f}_i$ in Groot's method consists of four components, as

$$\mathbf{f}_i = \sum_{j \neq i}(\mathbf{F}^C_{ij} + \mathbf{F}^D_{ij} + \mathbf{F}^R_{ij} + \mathbf{F}^S_{ij}) \,. \tag{3}$$

In the right hand side of this equation, the first three terms are the forces of the original DPD formulation [23, 24] to be considered within a certain radius $r_c$ under short-range cutoff. The crucial conservative force $\mathbf{F}^C_{ij}$ is a soft repulsion action as follows [25]

$$\mathbf{F}^C_{ij} = \begin{cases} -a_{ij}(1 - r_{ij})\mathbf{n}_{ij} & r_{ij} < 1 \\ 0 & r_{ij} \geq 1 \end{cases}, \tag{4}$$

where $a_{ij}$ is the maximum repulsion force between particles $i$ and $j$. The associated definitions are $\mathbf{r}_{ij} = \mathbf{r}_j - \mathbf{r}_i$, $r_{ij} = |\mathbf{r}_{ij}|$, and $\mathbf{n}_{ij} = \mathbf{r}_{ij}/|\mathbf{r}_{ij}|$. The repulsion parameter $a_{ij}$ between particles of different types corresponds to the mutual solubility, and is related to the Flory-Huggins χ parameter as [25]

$$a_{ij} = a_{ii} + 3.27\chi_{ij} \,. \tag{5}$$

In Eq. (3), the dissipative force $\mathbf{F}^D_{ij}$ [25] and the random force $\mathbf{F}^R_{ij}$ represent hydrodynamic drags and thermal noises of the Gaussian statistics, respectively.

The fourth term in the right hand side of Eq. (3) provides an additional spring force $\mathbf{F}^S_{ij}$ for directly bonded particles (or beads) in polymers [2, 25, 26]. For a certain connected particle pair $i$ and $j$, the corresponding harmonic force $\mathbf{F}^H_{ij}$ is given as

$$\mathbf{F}_{ij}^H = C(r_e - r_{ij})\mathbf{n}_{ij}, \tag{6}$$

where $C$ and $r_e$ are the force constant and equilibrium distance, respectively. This harmonic force is considered as the 1-2 type with direct connection. In Ref. [13], the Morse potential was utilized to express the 1-3 and 1-5 interactions required to describe the crucial hydrogen bonding in proteins. Here, we make a modification by introducing the absolute values for distances, in order to avoid the potential situation of strong repulsions at $r_{ij} < r_e$ region. The modified Morse force is then written as

$$\mathbf{F}_{ij}^M = 2K_M e^{-\alpha|r_{ij}-r_e|}\left(e^{-\alpha|r_{ij}-r_e|} - 1\right)\mathbf{n}_{ij}, \tag{7}$$

where $K_M$ and $\alpha$ are the well depth of Morse potential and well width, respectively. When the additional potentials for non-bonding interactions are incorporated, $\mathbf{F}_{ij}^S$ in Eq. (3) becomes the summation of harmonic and Morse contributions as

$$\mathbf{F}_{ij}^S = \mathbf{F}_{ij}^H + \mathbf{F}_{ij}^M. \tag{8}$$

This is notably different from the conventional DPD framework by Groot [1, 25].

# Design concept of CAMUS

In a usual molecular dynamics (MD) simulation software, the Verlet neighbor list method [27] and the associated cell lists method [28–30] are used to reduce the amount of calculations of non-bonding near-distance interaction. The former method is based on a task list to compute possible particle - particle interactions within a certain threshold of distance, and this list is usually updated every time step of preset interval. In the latter method, a given simulation box is divided into smaller cells, and the list of cell pairs with interactions is constructed. It is necessary to update the cell list. Those two methods are frequently used in combination, and there have been many associated variants [31]. For example, the COGNAC code [10–12] was designed to do simulations of both usual atomistic MD and DPD, and thus the above-mentioned list methods were implemented.

There could be a potential problem in DPD simulations as follows. The time step $\Delta t$ in DPD can be longer by 5-10 times that used in MD; in particular, $\Delta t$ is 0.05 in dimensionless unit [25]. Additionally, DPD particles move with both dissipative forces and random forces (recall Eq. (3)). Those two factors of DPD could provide large displacements of particles per simulation step, relative to usual MD simulations. As a whole, the Verlet neighbor list [27] should be constructed at every step of DPD simulations, leading to a potential overhead. Thus, in our CAMUS, the construction of neighbor list is abandoned (or not implemented), but the cell list containing interacting particles is formed at every step. Note that the length of a cell is equal to that of the cutoff. Figure 1 illustrates the schematic flow of time evolution, where the parallelization of costly force calculations is indicated. Groot's scheme [25] was adopted for time evolution in CAMUS, as in the case of COGNAC [10–12].

Besides the avoidance of construction of a neighbor list, CAMUS has another notable feature. That is the flexibility to handle additional specific interactions (potentials and derived forces). The modified 1-3 and 1-5 Morse potentials of Eq. 7 are usable (pre-implemented). The actual force computation is done in subroutine "calc_force_bond" (as will be shown later). When needed, other forms of potential/force can be added through in this subroutine: both modification and re-compilation are easy. A couple of python scripts "gen_input.py" and "lib.py" assist the preparation of definition list for bonds and specific interactions. From a viewpoint of educations, even graduate students may modify CAMUS for their respective purposes, based on a compact structure of this DPD code.

CAMUS was written as a portable Fortran program to which optimized compilers and libraries are available on various platforms. The force computation of particle - particle interactions is demanding in time evolution (refer to Fig. 1), and the thread-based parallelization was made for this part under the OpenMP shared-memory environment.

# Results and discussion

### Performance and parallel efficiency

The performance test of CAMUS was made, in comparison with COGNAC (written in C++) [10–12] that had been parallelized with OpenMP threads. The reason for the choice of COGNAC was due to the commonality in both time evolution [25] and parallelization and also our accumulated experiences of its usage (for example in Ref. [32]).

Two single-node servers were employed for the performance test. The first one was equipped with two Intel Xeon E5-2640 CPUs (clock-rate 2.50 GHz, 6 cores), and the second one was of many-core type with Intel Xeon Phi 7290 CPU (Knights Landing generation, clock-rate 1.50 GHz, 72 cores, compact/cache mode imposed). The binaries of CAMUS and COGNAC were built with standard Intel compilers and libraries.

The cubic box of DPD simulation was set for the numbers of particles of 5000, 10000, 50000 and 100000, by keeping the same reduced density of 3. The $a_{ij}$ in Eq. (5) was set to 25. The acceleration efficiency $Acc$ (%) is defined as

$$Acc = 100 \times \frac{T_{1\,thread}}{T}, \quad (9)$$

where the denominator $T$ and the numerator $T_{1\,thread}$ indicate the time with the parallelization by OpenMP threads (the corresponding number of threads is denoted as $N_{threads}$) and the reference time computed by 1 thread (or without parallelization), respectively. The parallelization efficiency $Par$ (%) is thus given as

$$Par = 100 \times \frac{T_{1\,thread}}{T \times N_{threads}}. \quad (10)$$

Table 1 shows the performance of CAMUS and COGNAC on the Xeon server. In all the cases of particles (and threads), CAMUS is considerably faster than COGNAC in the computational time. The time increment against the increase of particles is almost linear or sublinear for CAMUS, but such a preferable scaling is not observed for COGNAC. The acceleration efficiency of CAMUS is slightly better than that of COGNAC as well. The difference in performance between CAMUS and COGNAC nonlinearly enlarges according to the increase of particles: timings with a single thread by CAMUS and COGNAC are 10 (194) ms and 16 (630) ms, respectively, for the case of 5000 (100000) particles. Fig. 2 plots the comparative timings between CAMUS and COGNAC, where the cases of single and double threads are shown. The sizable difference in scaling behavior should be attributed to the fact that the particle - particle interactions are calculated without the neighbor list after the cell division in CAMUS. In other words, the construction of a neighbor list could be costly in DPD simulations. Unfortunately, for both CAMUS and COGNAC, the parallel efficiency quickly drops after 4 threads even for the case of 100000 particles. If much more particles (say $10^7$ - $10^8$ particles) are involved in actual simulations, further parallelization with a domain partitioning (which was done in LAMMPS [15, 16], HOOMD-blue [17–19], and DL_MESO [22]) should be necessary: a hybrid approach of OpenMP (thread) and MPI (process) may be a promising recipe.

In Table 2, the results on the Xeon Phi server are listed, where timing itself is slower than that of Xeon shown in Table 1 per the same number of threads. Overall performance behavior of CAMUS relative to COGNAC is similar to the results in Table 1. Unfortunately, the dropping trend in parallel efficiencies is again observed when the number of threads increases: particularly, more than 16 threads are not efficient. The use of regular Xeon CPU is recommendable for DPD simulations at the present implementation of CAMUS.

## Formation of protein structure

Besides Ref. [13], there have been several papers in DPD simulations for protein models [33–36]. Such applications will increase in the future. The fundamental applicability of CAMUS to proteins is thus of interest.

First, the reproduction of α-helix formation of a small protein model was checked. The simulation condition was almost the same as that reported in Ref. [13]. The values of $a_{ij}$ in Eq. (5) were set as $a_{SS} = a_{WW} = 50$ and $a_{SW} = 55$, where subscripts "$S$" and "$W$" mean the skeletal (or main chain) particles and water particles, respectively. The simulation system consisted of 24000 particles in a simulation box of $20 \times 20 \times 20$ size: the number of "$S$" particles was 60. Fig. 3 shows the force computation part in subroutine "calc_force_bond" and the definitions of bonds needed for the α-helix formation. This simple definition list (Fig. 3b) is to be processed by a couple of python scripts (refer to the previous section). For the 1-2 harmonic bond in the protein model, the parameters of $C = 160$ and $r_e = 0.6$ were used. The additional 1-3 and 1-5 interactions in the protein model are illustrated in Fig. 4. The 1-3 interaction consisted of two different components:

(i) harmonic bond with setting of $C = 80$ and $r_e = 1.2$, (ii) the Morse bond described by $K_M = 12$, $\alpha = 8$, and $r_e = 0.9$. The distant 1-5 Morse bond had the parameters of $K_M = 12$, $\alpha = 8$, and $r_e = 0.6$. Those Morse interactions have been introduced to mimic hydrogen bonding [13]. The DPD parameters were chosen as $T = 200$, $k_B T = 1$, $\gamma = 4.2$, $\lambda = 0.65$ (where $\gamma$ and $\lambda$ were a couple of DPD algorithm parameters [1, 2]), and $\Delta t = 0.02$. The dimensionless density was again 3, and the number of steps was 10000. The Xeon server was used for this DPD simulation. As presented in Fig. 5, the formation of α-helix [13] was reproduced with CAMUS, indicating that the inclusion of 1-3 and 1-5 non-bonding interactions works well.

We also tried the formation of β-sheet structure. The model setting is summarized in Fig. 6. Only harmonic forces were used for simplicity, by considering that the Morse potential can be locally approximated with the harmonic potential around the well bottom. Note that the 1-4 interaction was introduced as well as the 1-3 and 1-5 interactions. The parameter set listed in Fig. 6 was set after ad hoc trials, and the condition of DPD was similar with the case of α-helix formation. As shown in Fig. 7, the structure of β-sheet was formed by the present model setting. The reproduction of both α-helix and β-sheet structures implied that DPD simulations with CAMUS have a promising applicability to proteins.

Finally, the performance of CAMUS is again addressed for the case with additional interactions. The simulation box used for the test in Table 1 was modified to contain the protein model represented with 60 skeletal ("$S$") particles (for α-helix formation), leading to the volume fraction of 3 %. The total numbers of particles as well as the reduced density were the same as the previous test of Table 1. The timing results on the Xeon server are given in Table 3. Comparison with the entries in Table 1 indicates that the inclusion of additional 1-3 and 1-5 interactions provides only small increments of computational time and also that a favorable scaling behavior is retained.

# Concluding remarks

In this paper, we reported the development of a portable DPD code named CAMUS. In this code, the costly construction of neighbor list is avoided, and the particle - particle interactions are directly computed with the cell list. The benchmark tests showed that CAMUS has a preferable linear scaling behavior with respect to the increase of particles. The structures of α-helix and β-sheet were successfully formed for protein models through the inclusion of additional potentials [13]. Recently, a non-empirical way to evaluate effective interaction parameters for DPD simulations has been developed by us [37], based on the fragment molecular orbital (FMO) calculations [38]. This new scheme would be applicable even to amino acid residues as the components of proteins, and related works have been underway. Lastly, it should be noted that CAMUS is now freely available by HD at the GitHub web site [39].

# Acknowledgement

This work was partly supported by Ministry of Education, Culture, Sports, Science and Technology (MEXT) as a social and scientific priority issue #6 (Accelerated Development of Innovative Clean Energy Systems) to be tackled by using post-K computer and also as a grant-in-aid (Kaken-hi) No. 16H04635. Additionally, the authors would thank Dr. Yuto Komeiji for fruitful comments on the manuscript.

Table 1. Performances of CAMUS and COGNAC on a server equipping two Intel Xeon E5-2640 CPUs. Time in ms per step.

|  |  | CAMUS |  |  | COGNAC |  |  |
| --- | --- | --- | --- | --- | --- | --- | --- |
| # of particles | Threads | Time | Acc | Par | Time | Acc | Par |
| 5000 | 1 | 10.1 | 100.0 | 100.0 | 16.3 | 100.0 | 100.0 |
|  | 2 | 6.1 | 165.4 | 82.7 | 10.6 | 153.8 | 76.9 |
|  | 4 | 3.7 | 274.4 | 68.6 | 6.9 | 236.3 | 59.1 |
|  | 8 | 3.0 | 340.8 | 42.6 | 4.7 | 346.8 | 43.4 |
|  | 12 | 2.8 | 365.3 | 30.4 | 4.7 | 346.8 | 28.9 |
|  | 16 | 2.9 | 350.4 | 21.9 | 4.4 | 370.4 | 23.2 |
| 10000 | 1 | 20.5 | 100.0 | 100.0 | 34.9 | 100.0 | 100.0 |
|  | 2 | 11.9 | 172.9 | 86.4 | 24.6 | 141.9 | 70.9 |
|  | 4 | 7.6 | 268.8 | 67.2 | 13.3 | 262.4 | 65.6 |
|  | 8 | 9.5 | 215.4 | 26.9 | 9.0 | 387.8 | 48.5 |
|  | 12 | 7.1 | 289.5 | 24.1 | 8.1 | 430.9 | 35.9 |
|  | 16 | 5.1 | 400.0 | 25.0 | 7.5 | 465.4 | 29.1 |
| 50000 | 1 | 99.0 | 100.0 | 100.0 | 236.9 | 100.0 | 100.0 |
|  | 2 | 56.6 | 174.8 | 87.4 | 143.6 | 165.0 | 82.5 |
|  | 4 | 35.4 | 279.8 | 70.0 | 90.9 | 260.6 | 65.2 |
|  | 8 | 26.3 | 377.0 | 47.1 | 56.2 | 421.5 | 52.7 |
|  | 12 | 24.8 | 399.0 | 33.2 | 51.3 | 461.8 | 38.5 |
|  | 16 | 27.0 | 367.2 | 23.0 | 43.7 | 542.1 | 33.9 |
| 100000 | 1 | 194.5 | 100.0 | 100.0 | 634.4 | 100.0 | 100.0 |
|  | 2 | 113.5 | 171.4 | 85.7 | 393.8 | 161.1 | 80.6 |
|  | 4 | 68.4 | 284.5 | 71.1 | 232.4 | 273.0 | 68.2 |
|  | 8 | 48.7 | 399.4 | 49.9 | 143.3 | 442.7 | 55.3 |
|  | 12 | 56.8 | 342.5 | 28.5 | 128.5 | 493.7 | 41.1 |
|  | 16 | 52.3 | 371.8 | 23.2 | 109.4 | 579.9 | 36.2 |

Table 2. Performances of CAMUS and COGNAC on a server equipping Intel Xeon phi 7290 CPU. Time in ms per step.

|  |  | CAMUS |  |  | COGNAC |  |  |
| --- | --- | --- | --- | --- | --- | --- | --- |
| # of particles | Threads | Time | Acc | Par | Time | Acc | Par |
| 5000 | 1 | 65.2 | 100.0 | 100.0 | 152.3 | 100.0 | 100.0 |
|  | 2 | 37.2 | 175.2 | 87.6 | 92.2 | 165.1 | 82.6 |
|  | 4 | 20.5 | 318.5 | 79.6 | 49.6 | 307.0 | 76.8 |
|  | 8 | 15.0 | 433.8 | 54.2 | 30.4 | 500.7 | 62.6 |
|  | 16 | 10.0 | 652.6 | 40.8 | 23.5 | 647.1 | 40.4 |
|  | 32 | 10.9 | 599.3 | 18.7 | 25.2 | 604.8 | 18.9 |
| 10000 | 1 | 132.9 | 100.0 | 100.0 | 345.3 | 100.0 | 100.0 |
|  | 2 | 71.1 | 187.0 | 93.5 | 187.8 | 183.9 | 92.0 |
|  | 4 | 43.8 | 303.5 | 75.9 | 111.7 | 309.3 | 77.3 |
|  | 8 | 25.2 | 527.8 | 66.0 | 65.4 | 527.7 | 66.0 |
|  | 16 | 17.6 | 756.4 | 47.3 | 44.0 | 785.6 | 49.1 |
|  | 32 | 19.4 | 686.6 | 21.5 | 39.8 | 868.3 | 27.1 |
| 50000 | 1 | 733.7 | 100.0 | 100.0 | 1856.4 | 100.0 | 100.0 |
|  | 2 | 414.7 | 176.9 | 88.5 | 1007.2 | 184.3 | 92.2 |
|  | 4 | 240.9 | 304.6 | 76.1 | 586.5 | 316.5 | 79.1 |
|  | 8 | 154.6 | 474.6 | 59.3 | 344.0 | 539.6 | 67.5 |
|  | 16 | 98.3 | 746.3 | 46.6 | 227.8 | 814.8 | 50.9 |
|  | 32 | 89.4 | 820.6 | 25.6 | 167.8 | 1106.6 | 34.6 |
| 100000 | 1 | 1440.1 | 100.0 | 100.0 | 3789.3 | 100.0 | 100.0 |
|  | 2 | 808.1 | 178.2 | 89.1 | 2110.0 | 179.6 | 89.8 |
|  | 4 | 455.1 | 316.4 | 79.1 | 1181.7 | 320.7 | 80.2 |
|  | 8 | 261.5 | 550.8 | 68.8 | 684.7 | 553.5 | 69.2 |
|  | 16 | 184.6 | 780.1 | 48.8 | 437.9 | 865.3 | 54.1 |
|  | 32 | 175.6 | 820.0 | 25.6 | 314.6 | 1204.5 | 37.6 |

Table 3. Performance of CAMUS with 1-3 and 1-5 potentials (see text) on a server equipping two Intel Xeon E5-2640 CPUs. Time in ms per step.

| # of particles | Threads | Time | Acc | Par |
|---|---|---|---|---|
| 5000 | 1 | 10.2 | 100.0 | 100.0 |
|  | 2 | 6.2 | 165.9 | 82.9 |
|  | 4 | 4.9 | 211.0 | 52.8 |
|  | 8 | 3.0 | 346.4 | 43.3 |
|  | 12 | 2.8 | 372.1 | 31.0 |
|  | 16 | 2.9 | 352.6 | 22.0 |
| 10000 | 1 | 20.8 | 100.0 | 100.0 |
|  | 2 | 12.1 | 172.8 | 86.4 |
|  | 4 | 7.8 | 265.6 | 66.4 |
|  | 8 | 5.2 | 402.2 | 50.3 |
|  | 12 | 7.1 | 295.1 | 24.6 |
|  | 16 | 5.2 | 404.5 | 25.3 |
| 50000 | 1 | 104.9 | 100.0 | 100.0 |
|  | 2 | 60.9 | 172.1 | 86.1 |
|  | 4 | 37.4 | 280.8 | 70.2 |
|  | 8 | 25.9 | 404.5 | 50.6 |
|  | 12 | 26.3 | 398.7 | 33.2 |
|  | 16 | 27.9 | 375.8 | 23.5 |
| 100000 | 1 | 194.7 | 100.0 | 100.0 |
|  | 2 | 120.5 | 161.6 | 80.8 |
|  | 4 | 74.5 | 261.3 | 65.3 |
|  | 8 | 49.4 | 394.5 | 49.3 |
|  | 12 | 57.7 | 337.8 | 28.1 |
|  | 16 | 54.3 | 358.6 | 22.4 |

```
Time evolution:
  ! 1. update position
  position = position + dt*velocity + 1/2*dt2*force
  call boundary_check(position)

  ! 2. update velocity
  velocity2 = velocity + 0.65*dt*force

  ! 3. cell division
  call cell_division

  ! 4. calculate force
  Loop over cell number ! parallelized for cell index
    force2 =  calc_force(position, velocity2)
  End of loop over cell number

  ! 5. update velocity
  velocity = velocity + 1/2*dt*(force + force2)
```

Fig. 1. Schematic flow for time evolution in CAMUS. "dt" means the time step.

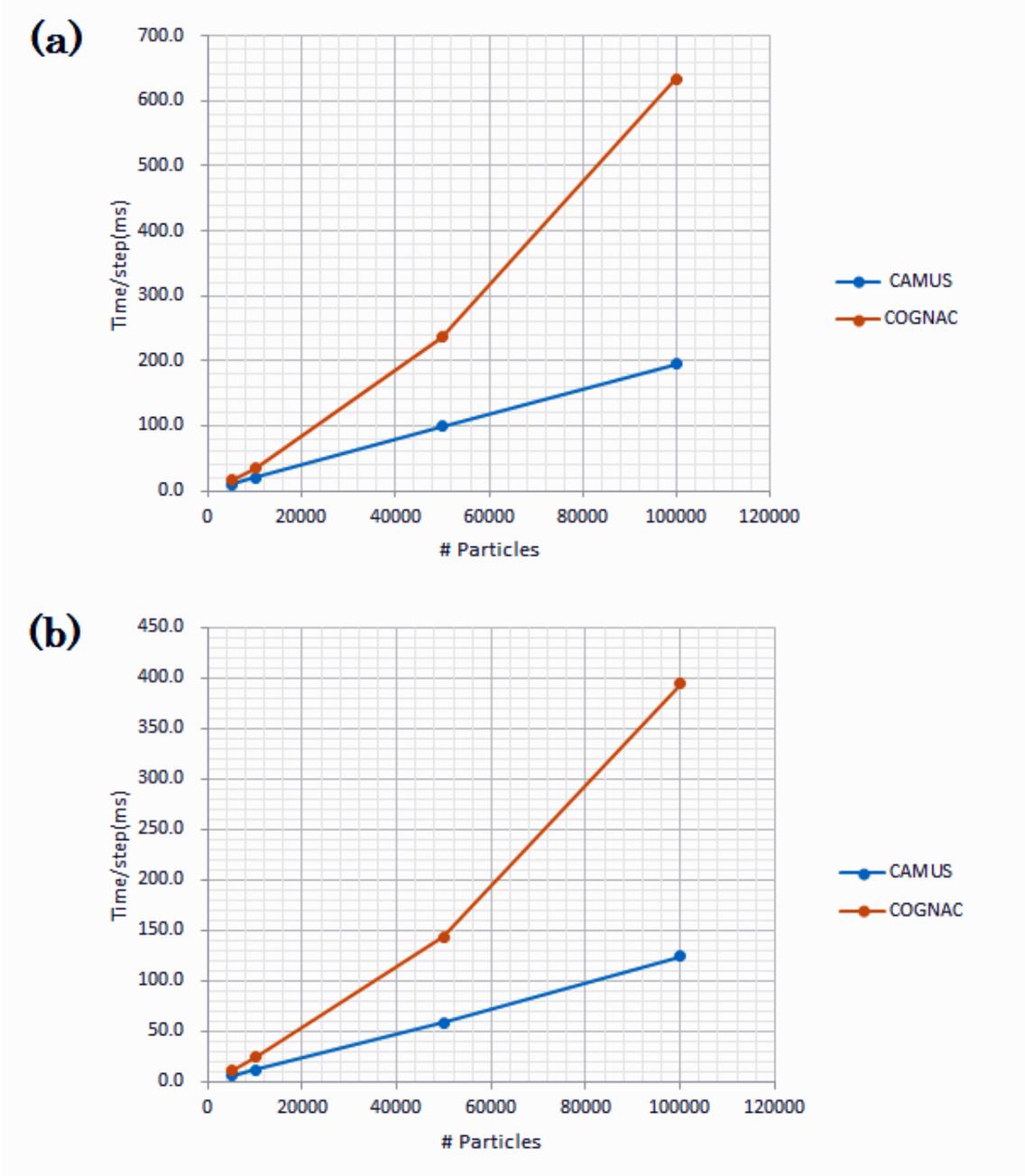

Fig. 2. Comparative timing plots between CAMUS and COGNAC. Timing data taken from Table 1 (on a server equipping Intel Xeon E5-2640 CPUs). (a) Results of 1 thread, (b) Results of 2 threads.

```
(a)   if( bond_list(now_bond)%bond_type .eq. 1 )then     ! Harmonic bond
        f(1:3) =  bond_list(now_bond)%const*dx*dr0
        Ebond = Ebond + 0.5*bond_list(now_bond)%const*dx*dx
      elseif( bond_list(now_bond)%bond_type .eq. 2)then   ! Morse bond
        exppart = exp(-bond_list(now_bond)%const2*abs(dx))
        expminusone = exppart - 1.0
        const_exp = bond_list(now_bond)%const*expminusone
        f(1:3) = 2.0d0*const_exp*bond_list(now_bond)%const2*exppart*dr0
        Ebond = Ebond + const_exp*expminusone
      endif

(b)   beads = 60
      bond12  = [ [1, i , i +1, 0.6, 160.0, 0] for  i  in range(beads-1)]
      bond13  = [ [1, i , i +2, 1.2,  80.0, 0] for  i  in range(beads-2)]
      bond13m = [ [2, i , i +2, 0.9,  12.0, 8] for  i  in range(beads-2)]
      bond15  = [ [2, i , i +4, 0.6,  12.0, 8] for  i  in range(beads-4)]

      monomers = [
      {'name':'water','particle': ['W'], 'move' : [1],'bond'  : []},
      {'name':'S60',
      'particle':['A' for i in range(beads)],
      'move':[1 for i in range(beads)],
      'bond':bond12+bond13+bond13m+bond15}
      ]
```

Fig. 3. Processing of α-helix formation (see text). (a) Force computation part of subroutine "calc_force_bond" in CAMUS, (b) Definitions for 1-2, 1-3 and 1-5 bonds (symbols of "1" and "2" in the square bracket of bond definition correspond to the harmonic and Morse types, respectively) as an input-data deck.

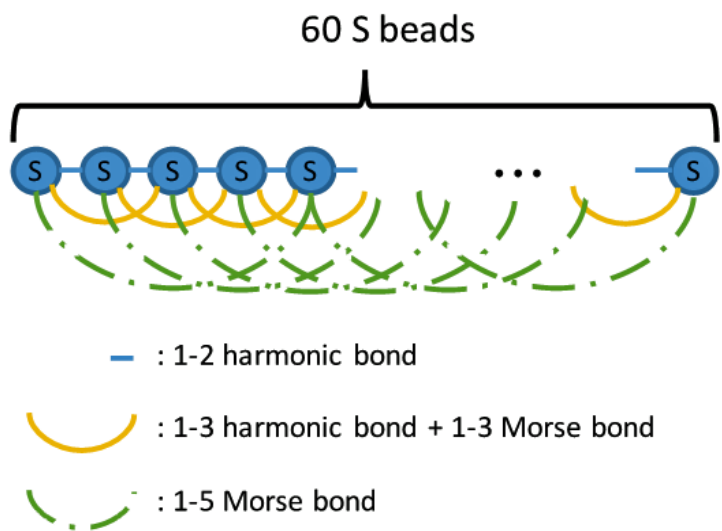

Fig. 4. Model setting for α-helix formation (see text). Refer also to Ref. [13].

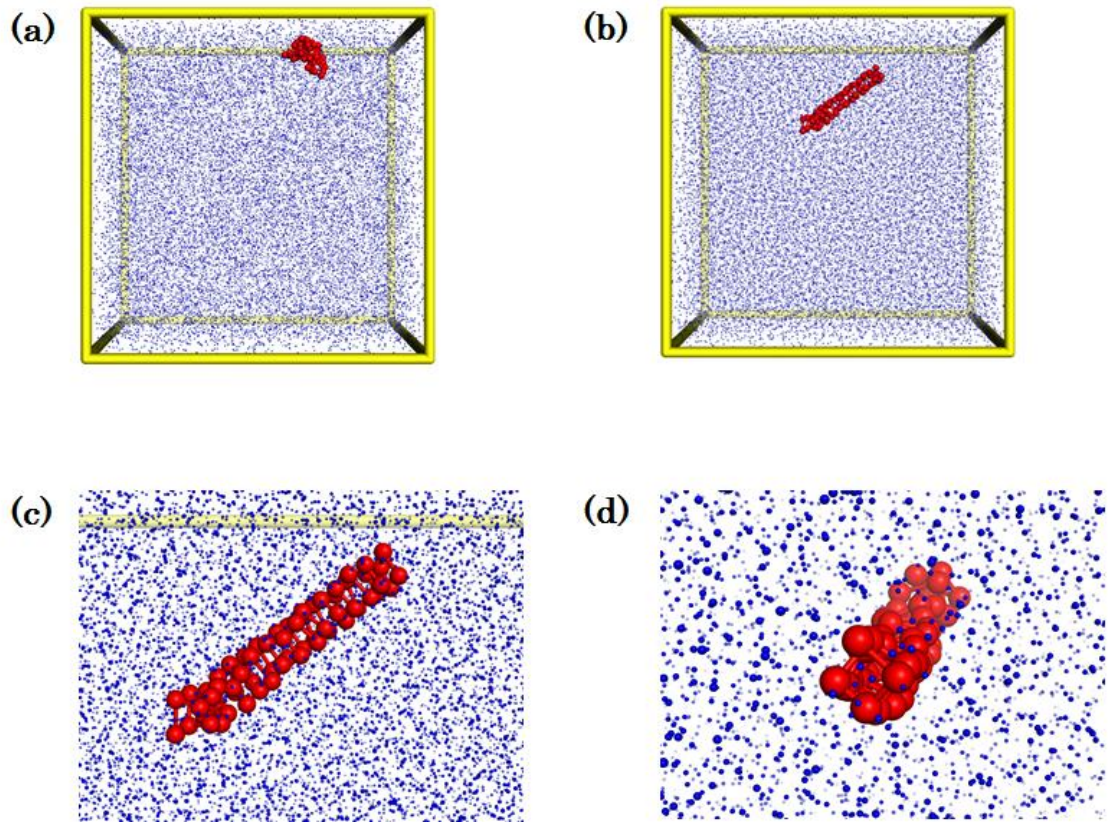

Fig. 5. Snapshot in DPD simulation for α-helix formation. Red and blue balls represent "$S$" and "$W$" particles, respectively (see text). (a) Initial structure, (b) Structure at $T = 200$, (c) Zoomed snapshot of α-helix structure, (d) Top view of α-helix structure.

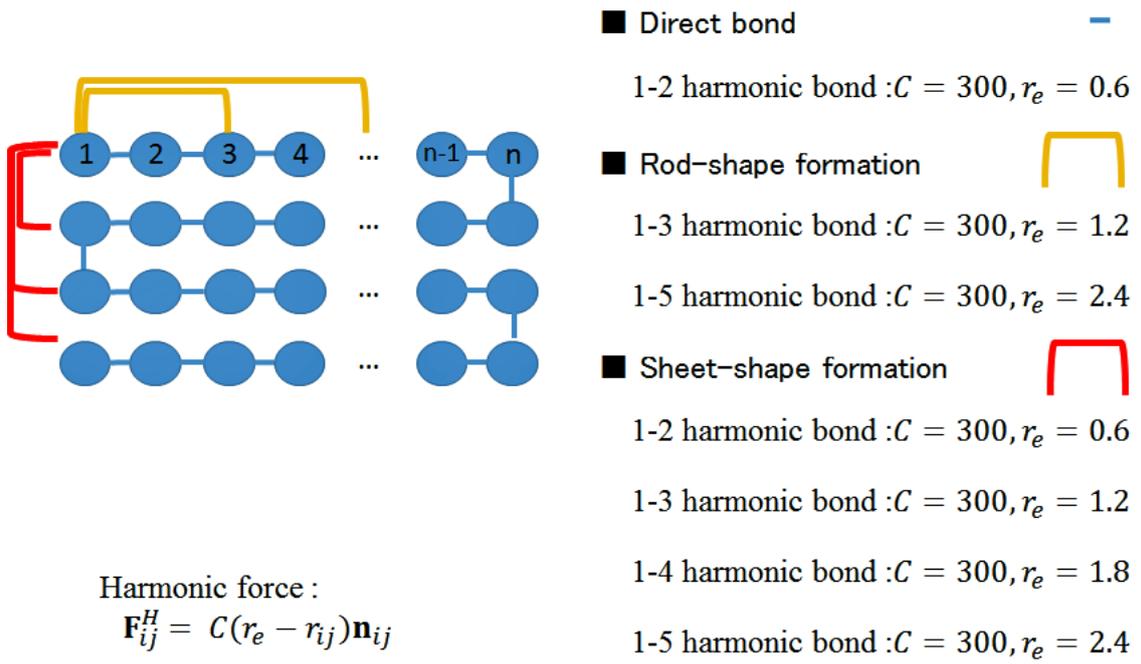

Fig. 6. Model setting for β-sheet formation (see text). Length of rod (n) was set to 10.

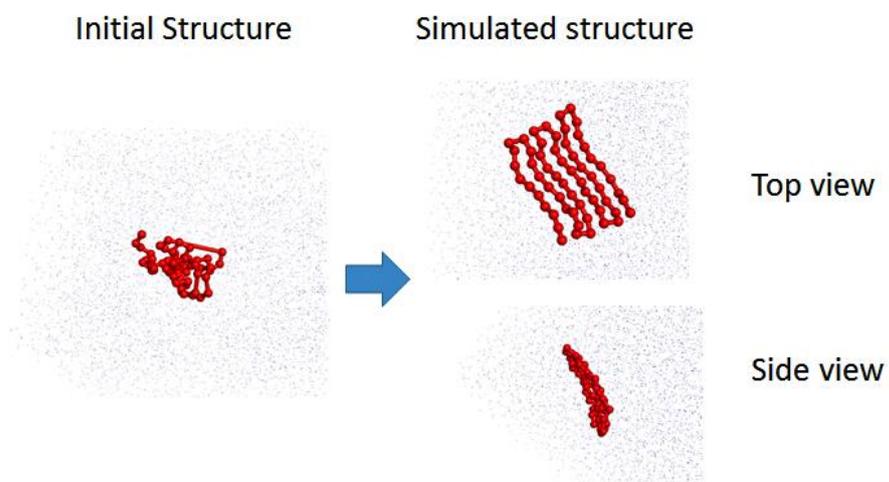

Fig. 7. Results of DPD simulation for β-sheet formation. Red and blue balls represent "$S$" and "$W$" particles, respectively (see text).